\documentclass[useAMS,usenatbib]{mn2e}
\usepackage[dvips]{graphicx}

\title[A survey for pulsars in EGRET error boxes]
      {A survey for pulsars in EGRET error boxes}
\author[D.~J.~Champion et al.]
{D.~J.~Champion,\thanks{Email: champion@jb.man.ac.uk} M.~A.~McLaughlin and D.~R.~Lorimer\\
University of Manchester, Jodrell Bank Observatory, Macclesfield, Cheshire, SK11 9DL, UK\\
}

\begin{document}

\date{\today}

\pagerange{\pageref{firstpage}--\pageref{lastpage}} \pubyear{2005}

\maketitle

\label{firstpage}

\begin{abstract}
As part of an effort to associate the unidentified EGRET $\gamma$-ray sources with pulsars, error boxes for 19 sources were searched using Arecibo at 327-MHz. The sources were chosen to be out-of-plane and possibly associated with the Gould belt, a nearby starburst region with an enhanced production rate of core-collapse supernovae. The search revealed one new 597-ms pulsar, J2243+1518, within the error box of the EGRET source 3EG~J2243+1509. The spin-down energy loss rate of the new pulsar is not nearly sufficient to power the $\gamma$-ray source and so the pulsar is very unlikely to be associated. Simulations we have carried out show that any pulsars at Gould belt distances should have been detected by the survey. This suggests that either the EGRET sources associated with the Gould belt are not pulsars, or that the minimum of the pulsar luminosity function is lower than the $\sim1.5$~mJy~kpc$^2$ inferred from the population of normal pulsars.
\end{abstract}


\section{Introduction}
One long-standing mystery in astrophysics is the origin of the unidentified $\gamma$-ray sources. The Energetic Gamma Ray Experiment Telescope (EGRET) produced a catalogue of 271 $\gamma$-ray sources at 100 MeV to 10 GeV energies (Hartman et al. 1999)\nocite{hbb+99}. These included 169 sources not identifiable with either pulsars or blazars, the two known classes of $\gamma$-ray source. While some promising identifications have been suggested since that time, the majority of these sources remain unidentified. As many as 40 of these unidentified sources have been shown to be statistically associated with the so-called ``Gould belt'', a dense region of atomic and molecular gas containing young stars \citep{gmb+00, gre00a, hz01}. The Gould belt is $\sim$~100~pc from the Sun at its nearest point and covers a region of radius $\sim$~300~pc at an angle of nearly 20$^\circ$ from the Galactic plane. Its nature as a starburst region means it produces core-collapse supernovae at an enhanced rate of 75 to 95~Myr$^{-1}$~kpc~$^{-2}$, $\sim$~4~times the Galactic rate. Since core-collapse supernovae are the most likely mechanism for pulsar production, this suggests that the belt should contain a sizable population of young pulsars \citep{gre00a}.

With the possible exception of one millisecond pulsar \citep{khv+00b}, all of the 6--8 EGRET pulsars from which $\gamma$-ray pulsations have been detected are relatively young and energetic, with characteristic ages less than 500~kyr and spin-down energy loss rates greater than $3\times10^{34}$~ergs~s$^{-1}$. Several convincing associations between newly discovered pulsars and unidentified EGRET sources have also been made (e.g. Kramer et al. 2003 \nocite{kbm+03} and Roberts 2005\nocite{rob05}). These pulsars are also young and energetic. Therefore, most previous surveys of EGRET error boxes for pulsar conterparts have concentrated on those sources close to the Galactic plane, where younger pulsars are likely to be found.

The combination of the Gould belt's potential to contain a population of young pulsars and its statistical association with the unidentified EGRET sources has prompted investigations of EGRET error boxes away from the plane. One previous search of mid-latitude error boxes resulted in no convincing Gould belt pulsar detections or EGRET pulsar counterparts \citep{rrh+04}. In this paper, we report on a search of 19 EGRET error boxes using Arecibo at a radio frequency of 327 MHz. The sources were chosen to be out of the plane and possibly associated with the Gould belt. This survey resulted in the discovery of PSR J2243+1518, a 597-ms pulsar with a spin-down age of 84~Myr. In Section \ref{Obs} we summarize our survey observations and data processing. The timing of PSR~J2243+1518 is described in Section \ref{Res}. We discuss the possible association of the new pulsar with 3EG~2243+1509 and test the hypothesis that pulsars in the Gould belt contribute to the EGRET sources in Section \ref{Disc}. Finally, in Section \ref{conc} we summarize our main conclusions and briefly review other possible origins for the EGRET pulsars.

\section{Observations and Analysis}
\label{Obs}
In August 2003 the Arecibo radio telescope was used to take the data presented here. The high Galactic latitudes ensured that the dispersion measure (DM) of any newly discovered pulsar would be small (less than $\sim 200$~pc~cm$^{-3}$ according to the Cordes \& Lazio 2002 model for Galactic free electron density\nocite{cl02}). This allowed us to observe at low frequencies, where pulsar emission is generally strongest. The newly commissioned 327-MHz receiver\footnote{see http://www.naic.edu~astro} system in the Gregorian dome was used in combination with the Wideband Arecibo Pulsar Processor (WAPP), a fast-dump digital correlator with a programmable number of lags and time resolution \citep{dsh00}. In this case 512~lags and a sample time of 125~$\mu$s were used. The Gregorian dome 327-MHz system has a gain of 11~K Jy$^{-1}$, a bandwidth of 25~MHz and a system temperature of 113~K. Each pointing was integrated for 260~seconds, giving an average limiting detectable flux density at 327~MHz of 0.4~mJy to pulsars with a period of $\sim$30~ms, pulse duty cycle of 10\% and a DM of 40~cm$^{-3}$~pc. For a typical pulsar spectral index of --1.6 \citep{lylg95}, the equivalent survey sensitivity at 430~MHz, a frequency more commonly used for pulsar searches at Arecibo, is $S_{\rm min, 430}=0.3$~mJy.

We selected a sample of 19 unidentified EGRET sources from the third EGRET catalogue with Galactic latitudes $|{\rm b}|>5^{\circ}$ and with declinations in the range $-2^{\circ}<\delta<38^{\circ}$ (i.e. visible with Arecibo). These sources are listed in Table~\ref{Sources}. The  1-$\sigma$ error boxes were gridded in an hexagonal pattern with beam centres separated by the beam width. The low frequency had the additional advantage of having a large beam size (14~$'$ at 327~MHz) and requiring fewer pointings to cover the error boxes. In total, 224 individual pointings were observed, with the coverage of the 1-$\sigma$ error boxes for all but five sources complete.

\begin{table*}
\caption{The selected unidentified EGRET sources accessible to Arecibo with $|$b$|>5^\circ$.\label{Sources}}
\begin{center}
\begin{tabular}{l c c c c c c c}
\hline
Name & $l$ & $b$& Error ellipse  & Mean error ellipse    & Number of       & $S_{\rm min}$ & Offset from \\
    3EG & (deg)       & (deg)      & diameter (deg) & area (deg$^{2}$) & pointings$^{a}$ & (mJy$^{c}$)   & Gould belt (deg)\\
\hline
J0215+1123 & 153.7 & $-$46.3 & 1.06 & 0.88 & 19           & 0.43 & 30.6\\
J0239+2815 & 150.2 & $-$28.8 & 0.47 & 0.17 & 7            & 0.44 & 14.2\\
J0348+3510 & 159.0 & $-$15.0 & 0.74 & 0.43 & 7            & 0.44 & 1.2\\
J0416+3650 & 162.2 & $-$09.9 & 0.63 & 0.31 & 4(7)$^{b}$   & 0.47 & 6.7\\
J0423+1707 & 178.4 & $-$22.1 & 0.77 & 0.47 & 13           & 0.45 & 3.0\\
J0426+1333 & 181.9 & $-$23.8 & 0.45 & 0.16 & 7            & 0.43 & 4.3\\
J0439+1555 & 181.9 & $-$19.9 & 0.92 & 0.66 & 18(19)$^{b}$ & 0.46 & 0.5\\
J0500+2529 & 177.1 & $-$10.2 & 0.36 & 0.10 & 1            & 0.45 & 8.7\\
J0520+2556 & 179.6 & $-$6.4 & 0.86 & 0.58 & 13           & 0.49 & 12.8\\
J0521+2147 & 183.0 & $-$8.4 & 0.45 & 0.16 & 5(7)$^{b}$   & 0.49 & 11.3\\
J1222+2315 & 241.8 &   +82.3 & 0.82 & 0.53 & 13           & 0.39 & 92.0\\
J1323+2200 & 359.3 &   +81.1 & 0.47 & 0.17 & 3(7)$^{b}$   & 0.40 & 61.8\\
J1347+2932 & 47.3 &   +77.5 & 0.95 & 0.71 & 19           & 0.40 & 59.8\\
J1822+1641 & 44.8 &   +13.8 & 0.77 & 0.47 & 13           & 0.53 & 3.5\\
J1824+3441 & 62.4 &   +20.1 & 0.82 & 0.53 & 13           & 0.44 & 6.4\\
J1825+2854 & 56.7 &   +18.0 & 0.97 & 0.74 & 19           & 0.47 & 3.0\\
J2243+1509 & 82.7 & $-$37.4 & 1.04 & 0.85 & 19           & 0.44 & 42.7\\
J2248+1745 & 86.0 & $-$36.1 & 0.94 & 0.69 & 19           & 0.41 & 40.4\\
J2352+3752 & 110.2 & $-$23.5 & 0.94 & 0.69 & 8(19)$^{b}$  & 0.42 & 20.5\\
\hline
\end{tabular}
\end{center}
In total, 224 independent beam positions were observed, corresponding to a total area of 9.6~deg$^2$.\\
$^{a}$The number of pointings taken to cover the error ellipse using the Arecibo telescope at 327 MHz.\\
$^{b}$Coverage incomplete due to time constraints. Number of pointings required appears in parentheses.\\
$^{c}$For a 30-ms pulsar with a pulse width of 3~ms, DM of 40~pc~cm$^{-3}$ and minimum S/N of 8.
\end{table*}

The observations were analysed at the Jodrell Bank Observatory using a 182-processor Beowulf cluster (COBRA, Joshi et al. 2003\nocite{jlk+03}). A suite of data analysis tools \citep{SIGPROC} was used in combination with scripts to keep each node continuously processing from the pool of data. Each pointing was dedispersed using 392 trial DMs between 0 and 491.2~pc~cm$^{-3}$. The upper DM limit is over twice as high as the expected DMs for any pulsars in this region
of the Galaxy. The resulting time series were then Fourier transformed and searched for periodic signals. Incoherent summing of the first 2, 4, 8 and 16 harmonics was used to increase sensitivity to narrow pulse profiles. Any resulting candidates with a signal-to-noise ratio (S/N) greater than 8 were folded for visual inspection. It took $\sim$~20,000 CPU~hours to complete the analysis of the 224 pointings.

\section{Results and Follow-Up Observations}
\label{Res}
\begin{figure}
\includegraphics[angle=0,scale=.33]{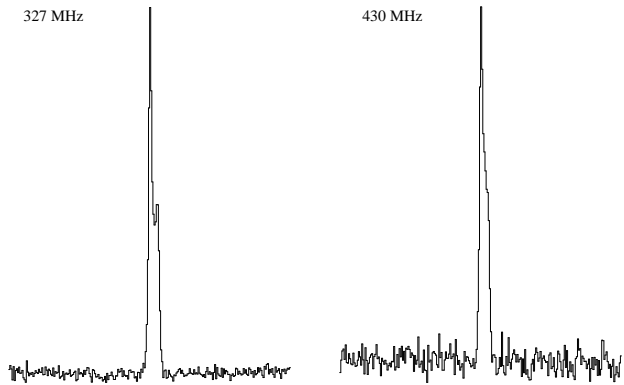}
\caption{The integrated pulse profile of PSR~J2243+1518 at 327 (left) and 430 (right) MHz.
These profiles are for $\sim$ 300 and $\sim$ 200 minutes of data at 327 and 430 MHz, respectively. Both profiles show full pulse phase over 256 bins. These profiles are freely available as part of the European Pulsar Network database (http://www.jb.man.ac.uk/$\sim$pulsar/Resources/epn).}
\label{Profiles}
\end{figure}

In the $\sim$~9.6~deg$^{2}$ of sky covered one promising candidate was found with $P=597$~ms, DM $\sim$~40~pc~cm$^{-3}$ and S/N~$\sim$~17. The candidate was re-observed and confirmed as a new pulsar, J2243+1518, in May 2004 using the same observing system with a 20-min integration. In Fig. \ref{Profiles} we present integrated pulse profiles at 327 and 430 MHz.

The pulsar was timed over the next 12 months at 327~MHz using the WAPPs and at 430~MHz using the Penn State Pulsar Machine (PSPM), a 128-channel analogue filterbank spectrometer which samples the incoming voltages from the telescope every 80~$\mu$s over a bandwidth of 7.68~MHz \citep{Cad97a}. The timing procedure used was identical to that described by Champion et al.~(2005)\nocite{clm+05}. In brief, using a preliminary ephemeris, the data were folded modulo the period and cross correlated with a high S/N template profile to obtain an accurate time of arrival (TOA) for each observation. The TOAs were then analysed with the TEMPO\footnote{http://pulsar.princeton.edu/tempo} software package to produce a phase-connected solution (where every rotation of the pulsar is accounted for) over the time span of the observations. This resulted in the ephemeris given in Table~\ref{PulsarPars}.

\begin{table*}
\caption{The measured and derived parameters for PSR~J2243+1518 based on 117 TOAs spanning 494 days.\label{PulsarPars}}
\begin{center}
\begin{tabular}{l l}
\hline
\multicolumn{2}{c}{Measured Parameters}\\
\hline
Right ascension (J2000) (h:m:s)                               & 22:43:09.768(4)\\
Declination (J2000) ($^\circ$:$'$:$''$)                       & 15:18:25.11(11)\\
Period (ms)                                                   & 596.799464458(4)\\
Period derivative ($\times 10^{-16}$)                         & 1.125(7)\\
Epoch of period (MJD)                                         & 53258\\
Dispersion measure (pc cm$^{-3}$)                             & 39.828(5)\\
327-MHz pulse width (10 per cent) (ms)                        & 30(2)\\
327-MHz pulse width (50 per cent) (ms)                        & 9(2)\\
327-MHz pulse width (eq) (ms)                                 & 15(2)\\
327-MHz flux density (mJy)                                    & 0.24(7)\\
430-MHz pulse width (10 per cent) (ms)                        & 29(2)\\
430-MHz pulse width (50 per cent) (ms)                        & 18(2)\\
430-MHz pulse width (eq) (ms)                                 & 14(2)\\
430-MHz flux density (mJy)                                    & 0.16(5)\\
RMS residual to fit ($\mu$s)                                  & 282\\
\hline
\multicolumn{2}{c}{Derived Parameters}\\
\hline
Galactic longitude (J2000) (deg)                              & 82.8\\
Galactic latitude (J2000) (deg)                               & $-$37.4\\
Distance (kpc)$^a$                                            & 3.5\\
Spin-down age (Myr)                                           & 84\\
Spin-down energy loss rate ($\times 10^{31}$ erg s$^{-1}$)  & 2.1\\
Surface dipole magnetic field strength ($\times 10^{11}$ G) & 2.6\\
\hline
\end{tabular}
\end{center}
Figures in parentheses are 1-$\sigma$ uncertainties in the least significant digits as calculated using TEMPO. The arrival times from which this ephemeris was derived is freely available on-line as part of the European Pulsar Network (EPN) database (http://www.jb.man.ac.uk/$\sim$pulsar/Resources/epn).\\
$^{a}$Inferred from the free electron density model of \cite{cl02}. Distance measurements using this model have a statistical error of $\sim$25\% although individual errors may be larger.
\end{table*}

\section{Discussion}
\label{Disc}
\subsection{PSR J2243+1518 and 3EG 2243+1509}
The spin parameters of PSR J2243+1518, listed in Table \ref{PulsarPars}, place it in the centre of the normal distribution of pulsars in the $P$-$\dot{P}$ diagram. Its $P$ and $\dot{P}$ imply a characteristic age of 84~Myr, a magnetic field strength of $B = 2.6\times10^{11}$~G and an energy loss rate of $\dot{E} = 2.1~\times10^{31}$~erg~s$^{-1}$. If this pulsar were responsible for the $\gamma$-ray emission detected by EGRET then this emission would have to be powered by its spin-down energy loss. The implied efficiency for conversion of spin-down energy into $\gamma$-rays is $\eta = L_{\gamma}/\dot{E} = 2600(4\pi f)(d/\rm 3.5~kpc)^{2}$, where $f$ is the beaming fraction (i.e. fraction of solid angle swept out by the $\gamma$-ray beam), a $\gamma$-ray photon index of --2 is assumed and the DM inferred distance of 3.5~kpc is used (see Table~\ref{PulsarPars}). If the pulsar were converting all of its spin-down energy into $\gamma$-rays, the distance would have to be $\sim$~0.07~kpc for it to be responsible for the $\gamma$-ray source. It is therefore highly unlikely that PSR~J2243+1518 is associated with 3EG~2243+1509.

The above arguments suggest that PSR~J2243+1518 is part of the background population of pulsars at intermediate/high Galactic latitudes. Based on the drift-scan surveys taken using the Arecibo telescope at 430~MHz, with a sensitivity of $S_{\rm min,~drift} \sim 1$~mJy we could expect to find such a pulsar every $\sim$~40~deg$^2$ \citep{mlc+04}. Assuming a spectral index of $-1.6$ to derive $S_{min,~ 430}$ and an isotropic population the number of pulsars expected by chance is $(9.6/40) \times (S_{\rm min,~430}/S_{\rm drift})^{-3/2} = 1.56$. It is, therefore, no surprise that we should find a pulsar which is not associated with the EGRET sources in this sample of error boxes.

\subsection{Testing the Gould Belt Hypothesis}
To test the hypothesis that the EGRET sources associated with the Gould belt are powered by pulsars, a Monte Carlo simulation was performed. All of the sources in Table~1 within 15$^{\circ}$ of the line of sight of the Gould belt (modelled as a plane inclined at 20$^\circ$ to the Galactic plane with the ascending node at $b=285^\circ$) were assumed to be associated with the Gould belt and the minimum luminosity of a detectable pulsar (based on our estimated $S_{\rm min}$) was calculated. A distance of 0.7~kpc was used, as this is the maximum distance of any part of the Gould belt \citep{gss+98, hz01}. A pulsar was then randomly picked from an assumed luminosity function with a lower bound of $L_{\rm min}=1.5$~mJy~kpc$^{2}$, upper bound of $L_{\rm max}=4.6\times10^{3}$~mJy~kpc$^{2}$ and a slope of d~$\log N$/d~$\log L = -1$ \citep{lml+98}. These bounds were scaled from the luminosity function derived by Lyne et al. to 327 MHz using a spectral index of --1.6. The flux density corresponding to this luminosity was then calculated and compared to $S_{\rm min}$ and the number of detectable model pulsars was recorded. In order to minimise statistical fluctuations in these results, this process was repeated 10,000 times and the results averaged.

Of the 19 sources we observed, 12 are within 15$^{\circ}$ of the line of sight of the Gould belt. The simulations always detected a pulsar counterpart to each of these sources; at a distance of only 0.7~kpc even a pulsar with a luminosity of $L_{\rm min}$ is detectable. Using the beaming model of \cite{tm98}, $\sim2/3$ of these pulsars will be beamed toward us, and given the coverage of the 1-$\sigma$ error boxes, 5.4 pulsars would be expected. Since no pulsars were found, it suggests that either the EGRET sources associated with the Gould belt are not pulsars or that the $L_{\rm min}$ for this population of pulsars is lower than $L_{\rm min}=1.5$~mJy~kpc$^{2}$ measured in the global population of normal pulsars. For example, an $L_{\rm min}<0.02$~mJy~kpc$^{2}$ is required before there is greater than a 50\% chance of not detecting any pulsars.

\section{Conclusion}
\label{conc}
Our search of 19 out-of-plane EGRET sources has resulted in the discovery of one new pulsar, J2243+1518. This new pulsar is very unlikely to be the source of the $\gamma$-ray emission due to the low  inferred spin-down energy loss rate. Our simulations show that, given the sensitivity of our search and the luminosity function of normal pulsars, it is unlikely that these error boxes harbour pulsars at distances that an association with the Gould belt would suggest.

Several authors have suggested other origins for the mid-latitude unidentified EGRET sources. A possible link between Galactic supernova remnants and 19 of the unidentified EGRET sources is described by \cite{tdr05}. The 19 positionally coincident sources are at low Galactic latitude but it is possible that the high supernovae production rate of the Gould belt could account for some of the mid-latitude sources. Low-mass microquasars are potential sources for the mid-latitude long-term variable sources \citep{gbr05}. However these sources are spatially distinct from the more stable sources in association with the Gould belt. X-ray and radio imaging of unidentified EGRET error boxes has resulted in the discovery of at least six pulsar wind nebulae \citep{rbg+05}. However, none of these are associated with Gould belt pulsars. Conversely, the sources could be radio-quiet pulsars like Geminga \citep{ggh+05} or off-beam $\gamma$-ray pulsars \cite{hz01}.

The best opportunity to determine the nature of the unidentified EGRET sources will come with next-generation $\gamma$-ray telescopes. The AGILE \citep{pcc+04} telescope will be able to detect pulsations from the brightest unidentified EGRET sources \citep{mc04}. GLAST \citep{GLAST}, however, will be able to detect pulsations from all unidentified EGRET sources, unambiguously determining their origin \citep{mc00}. Furthermore, its excellent angular resolution will allow deep radio searches for $\gamma$-ray source counterparts to be carried out.

\section*{Acknowledgements}
The Arecibo observatory, a facility of the National Astronomy and Ionosphere Center, is operated by Cornell University in a co-operative agreement with the National Science Foundation. We thank Paulo Freire for his advice on timing PSR~J2243+1518. We thank Alex Wolszczan for making the PSPM freely available for use at Arecibo. DJC is funded by the Particle Physics and Astrophysics Research Council. DRL is a University Research Fellow funded by the Royal Society.

\bibliographystyle{mn2e}
\bibliography{/psr/tex/bib/journals.bib,/psr/tex/bib/psrrefs.bib,thesisrefs.bib}

\label{lastpage}

\end{document}